\RequirePackage{lineno}

\documentclass[twocolumn,showpacs,superscriptaddress,preprintnumbers,amsmath,prc,amssymb]{revtex4}

\newcommand{\Heebar}{$^4\overline{\rm He}$}
\newcommand{\Hee}{$^4$He}
\newcommand{\He} {$^3$He}
\newcommand{\alphabar}{$\overline{\alpha}$}
\newcommand{\ebar}{$\overline{\rm e}$}
\newcommand{\pbar}{$\overline{\rm p}$}
\newcommand{\tbar}{$^3\overline{\rm H}$}
\newcommand{\db}{$\overline{\rm d}$}
\newcommand{\hypert}{$^3_{\Lambda}\rm H$}
\newcommand{\hypertbar}{$^3_{\bar{\Lambda}} \overline{\rm H}$}
\newcommand{\Hebar}{$^3\overline{\rm He}$}
\newcommand{\Libar}{$^6\overline{\rm Li}$}

\usepackage{lineno}

 \usepackage{epsfig}

\usepackage{amssymb}

\usepackage{graphicx}
\usepackage{dcolumn}
\usepackage{bm}
\usepackage{color}

\usepackage{CJK}

\begin{document}




\title{A brief review of antimatter production}

\author{Y. G. Ma \footnote{Author to whom correspondence should be addressed:
ygma@sinap.ac.cn} }
\affiliation{Shanghai Institute of Applied
Physics, Chinese Academy of Sciences, Shanghai 201800, China}

\author{ J. H. Chen}
 \affiliation{Shanghai Institute of Applied
Physics, Chinese Academy of Sciences, Shanghai 201800, China}

\author{L. Xue}
\affiliation{Shanghai Institute of Applied
Physics, Chinese Academy of Sciences, Shanghai 201800, China}
\affiliation{ University of
 the Chinese Academy of Sciences, Beijing 100080, China}


\begin{abstract} In this article, we present a brief review of  the
discoveries of kinds of antimatter particles, including positron
(\ebar), antiproton (\pbar), antideuteron (\db) and antihelium-3
(\Hebar). Special emphasis is put on the  discovery  of  the
antihypertriton(\hypertbar) and antihelium-4 nucleus (\Heebar, or
\alphabar) which were  reported by the RHIC-STAR experiment very
recently. In addition, brief discussions about the effort to
search for antinuclei in cosmic rays and study of the longtime
confinement of the simplest antimatter atom, antihydrogen are
also given. Moreover, the production mechanism of anti-light
nuclei is introduced. 
\end{abstract}

\pacs{ ~25.75.Dw, 13.85.Ni}

\maketitle

\begin{enumerate}

\item{Introduction}

\item{ Observation of positron}

\item{ Observations of antiproton, antideuteron and antihelium-3}

\item{ Observation of \hypertbar}

\item{ Observation of \Heebar}

\item{Effort to search for antinuclei in Cosmic
rays}

\item{ Trap of antihydrogen atoms}

\item{ Production mechanism of antimatter light nuclei}

\item{Conclusions and perspectives}
\end{enumerate}

\section{introduction}

The ideal of antimatter can be traced back to the end of 1890s,
when Schuster discussed the possibility of the existence of
antiatoms as well as antimatter solar system by hypothesis in his
letter to Nature magazine   \cite{Concept}. However, the modern
concept of antimatter is originated from the negative energy state
solution of a quantum-mechanical equation, which was proposed by
Dirac in 1928   \cite{Prediction}. Two years later, Chao found
that the absorption coefficient of hard $\gamma$-rays in heavy
elements was much larger than that was to expected from the
Klein-Nishima formula or any other \cite{Chao1,Chao2}. This
"abnormal" absorption is in fact due to the production of the pair of
electron and its anti-partner, so-called positron. Therefore
Chao's experiment is the first indirect observation of the first
anti-matter particle, namely positron, in the history. Another two
years later, Anderson observed positron   with a cloud chamber
\cite{Positron}. Antimatter nuclei such as \db~, \tbar~, \Hebar~
have been widely studied in both cosmic rays
\cite{AMS,Bess,Pamela} and accelerator experiments
\cite{Lederman,AntiHe3,Cork,AntiTritium,Phenixdbar,STARdHe3} for
the purposes of dark matter exploration and the study of manmade
matter such as quark gluon plasma (QGP) respectively, since the
observation of anti-proton (\pbar)  \cite{Chamberlain} in 1955.
The possibility of the anti-gravity behavior between matter and
antimatter has been discussed somewhere else  \cite{AntiG}. The
recent progress regard the observation of antihypertriton
(\hypertbar)  \cite{H3Lbar} and antihelium-4 (\Heebar, or
\alphabar)  \cite{He4bar} nucleus in high energy heavy ion
collisions \cite{Chen,Xue,Ma} reported by RHIC-STAR experiment as
well as the longtime confinement of antihydrogen atoms
\cite{ALPHA} based on an antiproton decelerator facility by ALPHA
collaboration have already created a lot of excitation in both of
the nuclear and particle physics community. All of the
measurements performed above have implications beyond the fields
of their own. Such as, the study of hypernucleus in heavy ion
collisions is essential for the understanding of the interaction
between nucleon and hyperon (YN interaction), which plays an
important role in the explanation of the structure of neutron
star. Furthermore, as we learned from heavy ion collisions, the
production rate for \Heebar~ produced by colliding the high energy
cosmic rays with interstellar materials is too low to be observed.
Even one \Heebar~ or heavier antinucleus  observed in the cosmic
rays should be a great hint of the existence of massive antimatter
in the Universe. Finally, the successful trap of antihydrogen
atoms leads to a precise test of the CPT symmetry law, as well as
a measurement of the gravitational effects between antimatter and
matter in the future.

In this article, we present a review of kinds of antimatter
particles experimentally, based on the time schedule of their
first time observations. The paper is arranged as follows, in
section 2, we will take a  look back to the discoveries of
positron and antiproton. In section 3, the first time observations
of antideuteron and antihelium-3 will be discussed. Section 4
presents a brief review of  the formation and observation of
\hypertbar~ through their secondary vertex reconstructions via
decay channel \hypertbar~$\rightarrow$\Hebar~+ $\pi^{+}$ with a
branch ratio of 25\% in high energy heavy ion collisions. In
Section 5, we have discussion of the particle identification of
\Heebar~nucleus by measuring their mass value directly with the
fully installed detector Time Of Flight (TOF) at RHIC-STAR. In
Section 6, we discuss the effort of searching for antinuclei in
cosmos. In Section 7, the longtime trap of antihydrogen atoms
performed by ALPHA experiment is introduced. In the last section,
we have a brief discussion on antimatter nuclei production
mechanism. Finally we give a simple summary and outlook.

\section{Discoveries of positron and antiproton}

In 1930, C. Y. Chao performed a few $\gamma$-ray scattering
experiments  on different elements \cite{Chao1,Chao2}.
$\gamma$-rays from Th C after being filtered through 2.7cm of Pb
were used as the primary beam. Al and Pb were chosen as the
represensitives of the light and the heavy elements. For Al the
$\gamma$-scattering is, within experimental error,  that predicted
by the Klien-Nishima formula which assumes that the removal of the
energy from the primary beam is entirely due to Compton scattering
of the extranuclear electrons. However, for Pb additional
scattering rays were observed. The wavelength and space
distribution of these are inconsistent with an extranuclear
scatterer \cite{Chao1,Chao2}. Later on, this abnormal absorption
was identified as the outcome of the process of  electron-positron
pair production. Therefore this experiment was also the first
experimental indiction of the first anti-matter particle,
positron, in the observation history for the antimatter.

-----------------------------------------------------------------------

\vspace{3mm}
\centerline{\psfig{figure=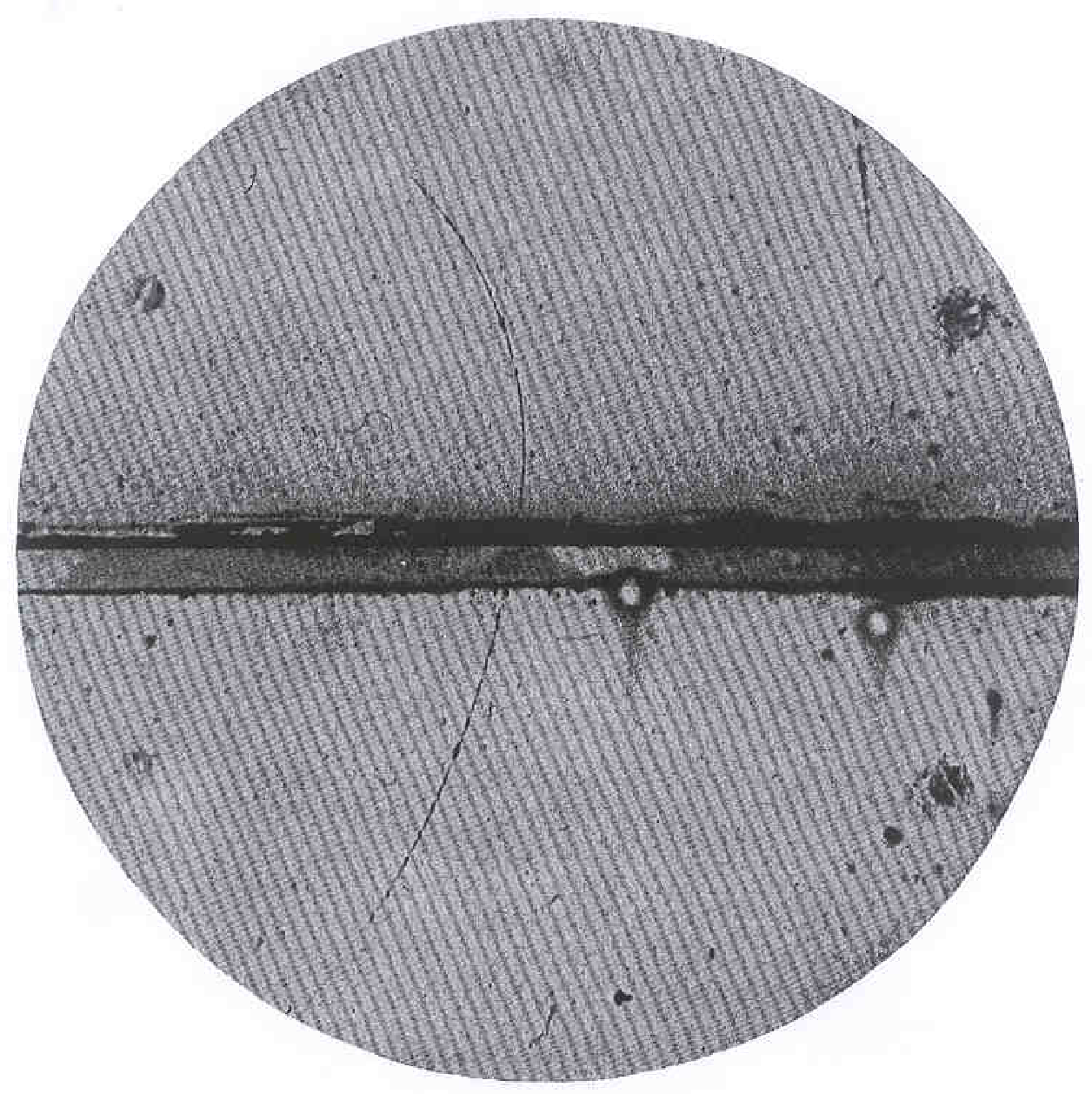,width=3.2in}\vspace{1mm}}
 {\baselineskip
10.5pt\renewcommand{\baselinestretch}{1.05}\footnotesize \noindent
{\bf Fig.~1}\quad A photograph for the discovery of positron from
reference    \cite{Positron}. The curvature stands for a comic ray
passing through a 6 mm lead plate with a reduction of its energy.
The length of the latter part can only be interpreted with the
appearance of antimatter electron. \label{positron} }\vspace{4mm}

Two years later,  C. D. Anderson identified 15 positive tracks by
photographing the cosmic rays with a vertical Wilson chamber in
August, 1932 \cite{Positron}. The unknown particles were
recognized as the predicted antimatter electron (positron) after
analyzing their energy loss, ionization, as well as their
curvatures in the chamber. Figure 1 shows that, a positron reduces
its energy by passing through a 6 mm lead plate in the cloud
chamber. The track length from the upper part of the cloud chamber
can only be interpreted with the observation of positron.

\section{ Observations of antiproton, antideuteron and antihelium-3}

Physicists had expanded their understanding of the natural world,
and  took into consideration that every particle should have their
antimatter partner after the discovery of positron. They were able
to expand their knowledge of antimatter with the development of
the technology of accelerators, while the development of
Time-of-Flight detector system played an important role in the
following identification of antimatter particles. In the year
1955, O. Chamberlain and E. Segr$\grave{\mathrm{e}}$ from
University of California reported their observation of antiprotons
based on the Bevatron facility  \cite{Chamberlain}. Antiprotons
were produced and scattered into the forward direction by
projecting a bunch of proton beam to the copper target at
Bevatron. By observing times of flight for antiprotons, it makes
more meaningful by the fact that the electronic gate
time is considerably longer than the spread of observed
antiproton flight times. The electronic equipment
accepts events that are within $\pm$6 millimicroseconds of
the right flight time for antiprotons, while the actual
antiproton traces recorded show a grouping of fight
times to $\pm$ 1 or 2 millimicroseconds. Figure 2(a) and 2(b) depicts
a histogram of meson flight times and  that of antiproton flight times, respectively.
 Accidental
coincidences account for many of the sweeps (about
$2/3$ of the sweeps) during the runs designed to detect
antiprotons. A histogram of the apparent flight times
of accidental coincidences is shown in Fig. 2(c). It will
be noticed that the accidental coincidences do not show
the close grouping of flight times characteristic of the
antiproton or meson flight times.
In total, 60 antiprotons were detected, with their mass
measured and found to be equal to protons within uncertainties
based on the detectors at Bevatron \cite{Chamberlain}.

\vspace{3mm}
\centerline{\psfig{figure=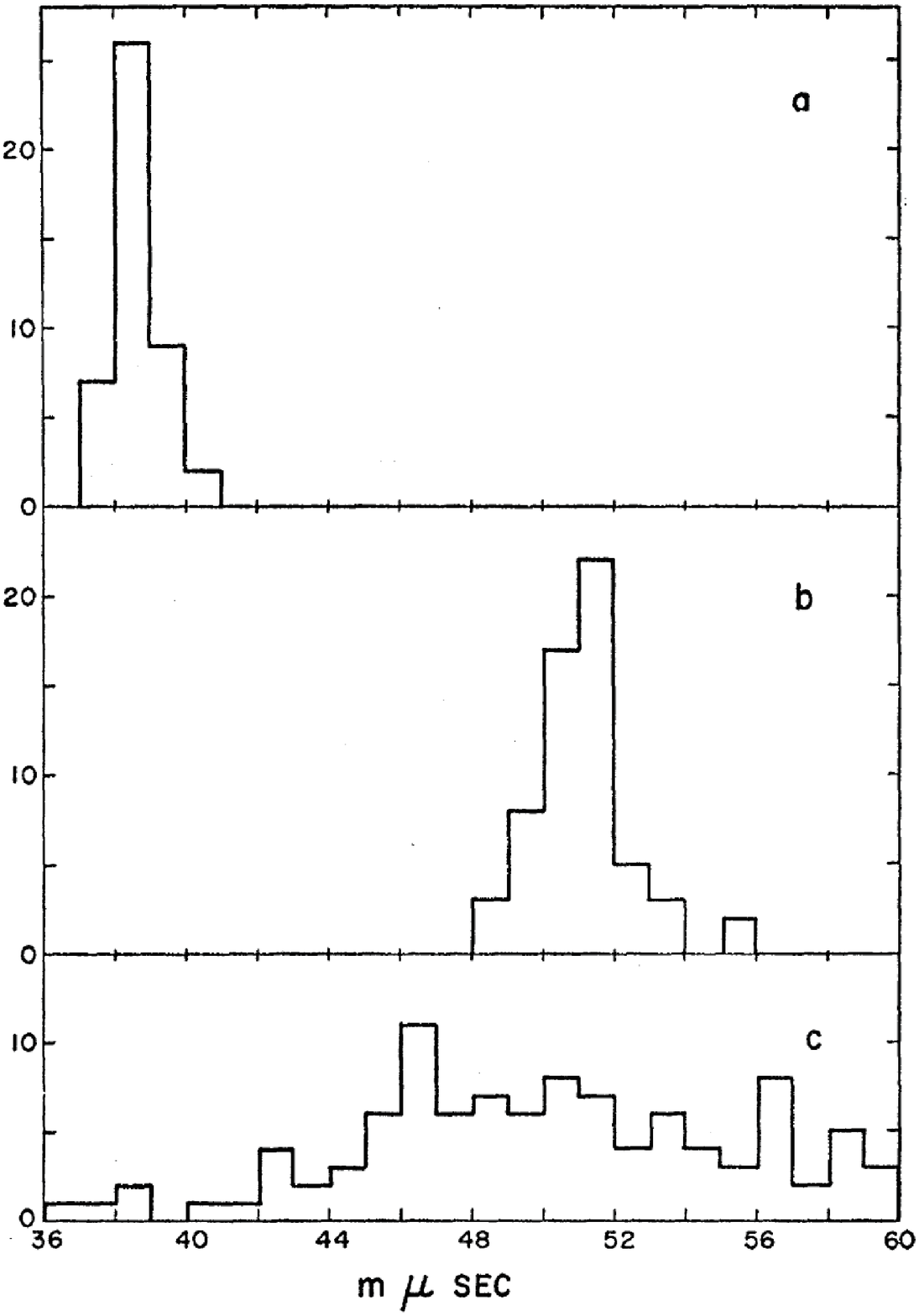,width=3.2in}\vspace{1mm}}
 {\baselineskip
10.5pt\renewcommand{\baselinestretch}{1.05}\footnotesize \noindent
{\bf Fig.~2}\quad (a) Histogram of meson times used for
calibration. (b) Histogram of antiproton flight times. (c)
Apparent  flight times of  a representative accidental
coincidences. Times of flight are in units of 10$^{-9}$ s.  The
ordinates show the number of events in each  10$^{-10}$ second
intervals. Taken from Ref.~\cite{Chamberlain}. \label{anti-p}
}\vspace{4mm}

Antideuteron was observed by Alternating Gradient Synchrotron
(AGS)   \cite{Lederman} at BNL and later confirmed by Proton
Synchrotron (PS)  \cite{Zichichi} at CERN in 1965, both of the
experiments were implemented by colliding protons with beryllium.
The data was collected with a set of time of flight system
S$_{1}$S$_{10}$ (210 ft) and S$_{2}$S$_{9}$ (170 ft) at AGS.
Figure 3(a), (b), (c) show the travel time of antideuterons
between detectors S$_{2}$ and S$_{9}$ at different momentum
region. The time to channel calibration is 0.56 ns per channel.
The peak positions from Figure 3(a),(b),(c) move from channel 21.5
at p = 6 GeV/c to channel 23.5 at 5.4 GeV/c, and to channel 26 at
5.0 GeV/c, corresponding to $\Delta \beta/\Delta p = (1.6\pm
0.4)\times 10^{-2} (MeV/c)^{-1}$. The calculated mass equals to
1.86 GeV/c via formula $d\beta/dp = \beta^{3}m^{2}/p^{3}$, and
interpreted to be antideuteron.

\vspace{3mm}
\centerline{\psfig{figure=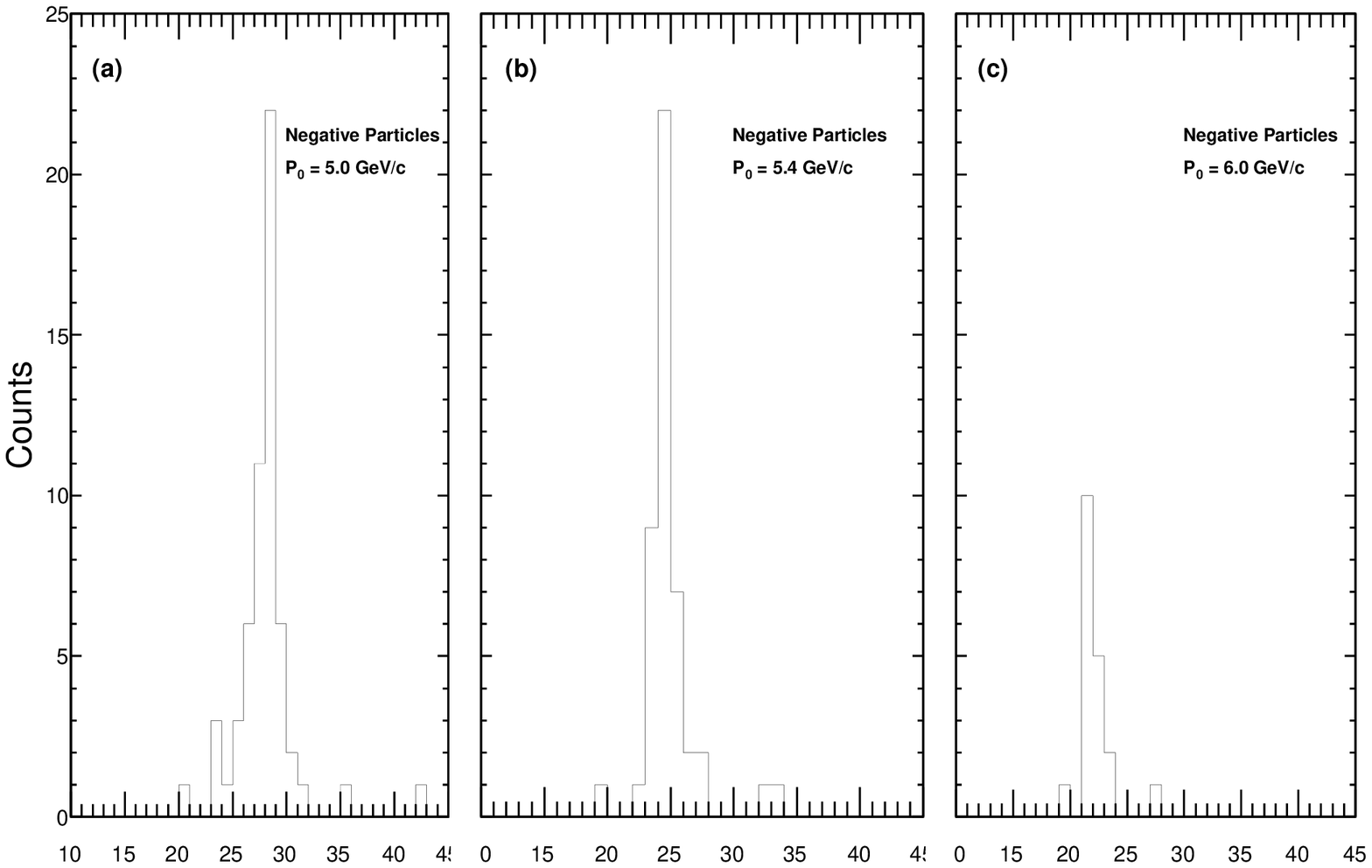,width=3.2in}\vspace{1mm}}
 {\baselineskip
10.5pt\renewcommand{\baselinestretch}{1.05}\footnotesize \noindent
{\bf Fig.~3}\quad The evidence for the observation of
antideuteron from Ref.~ \cite{Lederman}. (a) (b) (c) shows the
time-of-flight distribution for particles between counters S$_{2}$
and S$_{9}$ with 0.56 ns per channel. Taken from
Ref.~\cite{Lederman}. \label{antideuteron} }\vspace{4mm}

\vspace{3mm}
\centerline{\psfig{figure=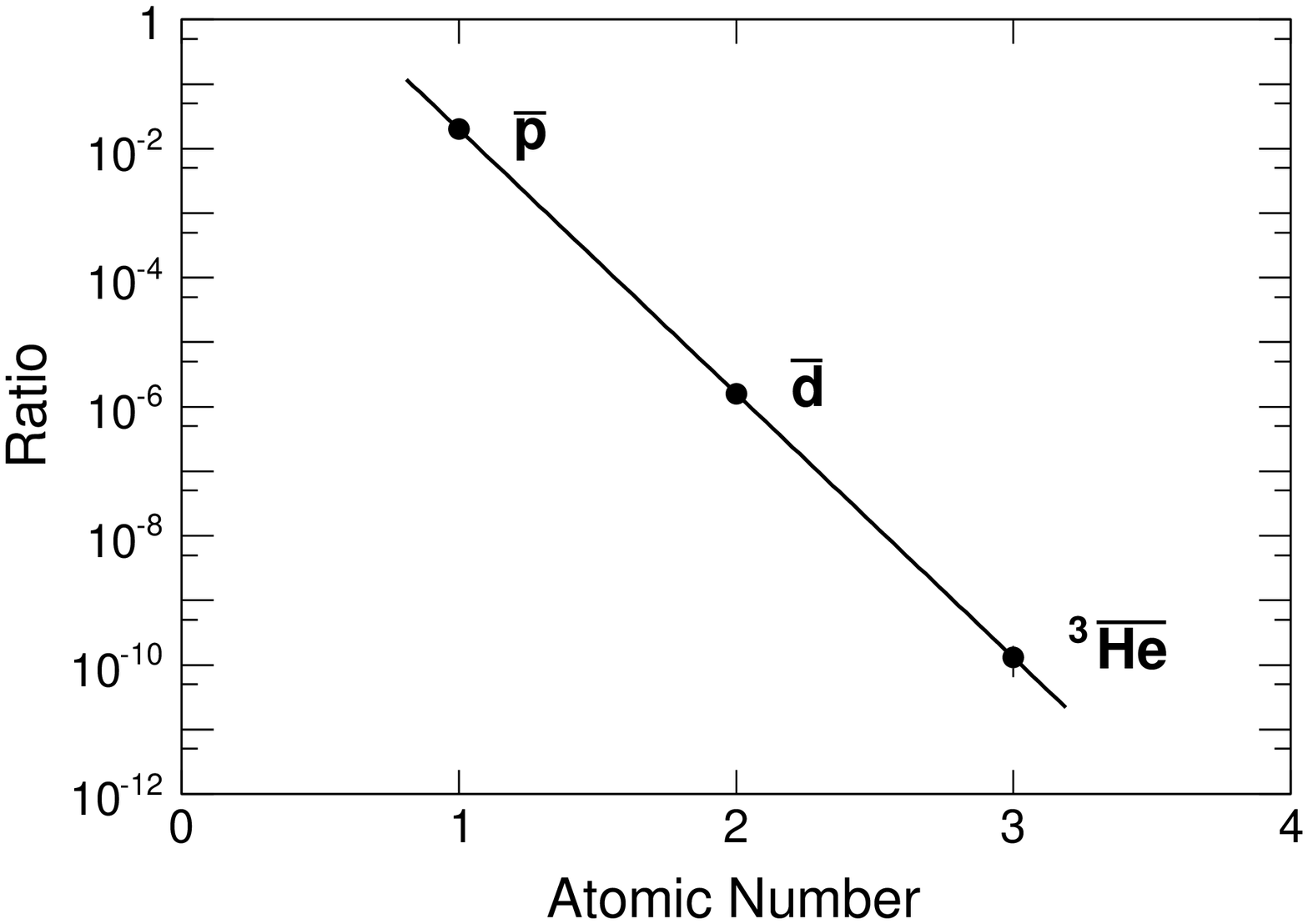,width=3.2in}\vspace{1mm}}
 {\baselineskip
10.5pt\renewcommand{\baselinestretch}{1.05}\footnotesize \noindent
{\bf Fig.~4}\quad Atomic number dependence of the relative
production of  antimatter nuclei with respect to pion with fixed
momentum at 20 GeV/c and $\theta$ equal to 27 mrad.  Taken from
Ref.~\cite{AntiHe3}. \label{antihelium3} }\vspace{4mm}

Experiments with antimatter nuclei heavier than antideuteron were
implemented at  the Institute of High Energy Physics, Russia. Five
antihelium-3  \cite{AntiHe3} were produced by colliding the 70GeV
proton beams with aluminum target. The calculated charge is
$(0.99\pm0.03)2e$, while the measured mass equals to
$(1.00\pm0.03)3m_{p}$. The differential production cross section
of antihelium-3 with momentum equals to 20 GeV/c at $\theta = 27
mrad$ was calculated and found to be $2.0\times10^{-35} cm^{2}/sr
\cdot GeV/c$ per aluminum nucleus. While the differential cross
section for $\pi^-$  is $1.5\times10^{-25} cm^{2}/sr \cdot GeV/c$
per aluminum nucleus. The ratio between differential cross section
of antihelium-3 and pion is $1.3\times10^{-10}$. Figure 4 shows an
atomic number dependence of the ratios. The ratios decrease by a
magnitude of four orders with each additional nucleon added. By
extrapolating the distribution to larger mass area, the relative
cross section of antihelium-4 is about $10^{-14}$.

\section{ Observation of \hypertbar}


Unlike the normal nuclei (anti-)nuclei, (anti-)hypernucleus
includes the (anti-)strange  quark degree of freedom, of which the
typical one is $\Lambda$-hypernucleus. The simplest hypernucleus
observed so far is hypertriton, which is composed of one neutron,
one proton and one $\Lambda$-hyperon. Hypernucleus provides an
ideal environment to study the hyperon-nucleon (YN) interaction,
responsible in part for the binding of hypernuclei, which is of
fundamental interest in nuclear physics and nuclear astrophysics.
In previous research history, quite lot hypernuclei have been
discovered, even for the observation of double-$\Lambda$
hypernucleus \cite{double-hyp}, but still no anti-hypernucleus was
observed until the STAR collaboration announced the first
anti-matter hypernucleus, i.e. \hypertbar \cite{H3Lbar}, in 2010.
The observation of \hypertbar~can be achieved by reconstructing
their secondary vertex via \hypertbar~$\rightarrow$\Hebar~+
$\pi^{+}$ \cite{H3Lbar,Chen}. The data used for analysis was
collected by STAR experiment at Relativistic Heavy Ion Collider
(RHIC), using the cylindrical Time Projection Chamber (TPC), which
is 4 meters in the diameter direction and 4.2 meters long in the
beamline direction \cite{TPC}. TPC is able to reconstruct charged
tracks between $\pm$1.8 units of pseudo-rapidity with full
azimuthal angle. The identification of tracks can be achieved by
correlating their ionization energy loss $\langle dE/dx \rangle$
in TPC with their magnetic rigidity. Figure 4C shows $\langle
dE/dx \rangle$ for negative tracks versus the magnetic rigidity.
The different bands stand for different kinds of particles. Figure
5 is the distribution of a new variable, z = $Ln(\langle dE/dx
\rangle/\langle dE/dx \rangle_{B})$,which is used to identify \He~
and \Hebar~, here $\langle dE/dx \rangle_{B}$ is the expected
value of $\langle dE/dx \rangle$. Approximately, a sample of 5810
\He~ and 2168 \Hebar~ were collected for the secondary vertex
reconstruction of \hypertbar.

Topological cuts including the distance between two daughter
tracks \Hebar~  and $\pi^{+}$ ($<$1cm), distance of closest
approach (DCA) between \hypertbar~ and primary vertex ($<$1cm),
decay length of \hypertbar ($>$2.4cm), and the DCA of $\pi$ track
($>$0.8cm), are employed to promote the signal to background
ratio. The invariant mass of \hypert~ and \hypertbar~ were
calculated base on the conservation of momentum and energy in the
decay process.  The results are shown in Figure 5A for \hypert~
and Figure 4B for \hypertbar~. The successfully reproduced
combinatorial background with a rotation strategy can be described
by double exponential function: $f(x) \propto exp[-(x/p_{1})] -
exp[-(x/p_{2})]$, where $x=m-m(^{3}He)-m(\pi)$, and $p_{1},p_{2}$
are the parameters. Finally, the signals are counted by
subtracting the double exponential background of \hypert~ and
\hypertbar~. In total, 157 $\pm$ 30 of \hypert~ and 70 $\pm$ 17 of
\hypertbar~ are observed.

\vspace{3mm}
\centerline{\psfig{figure=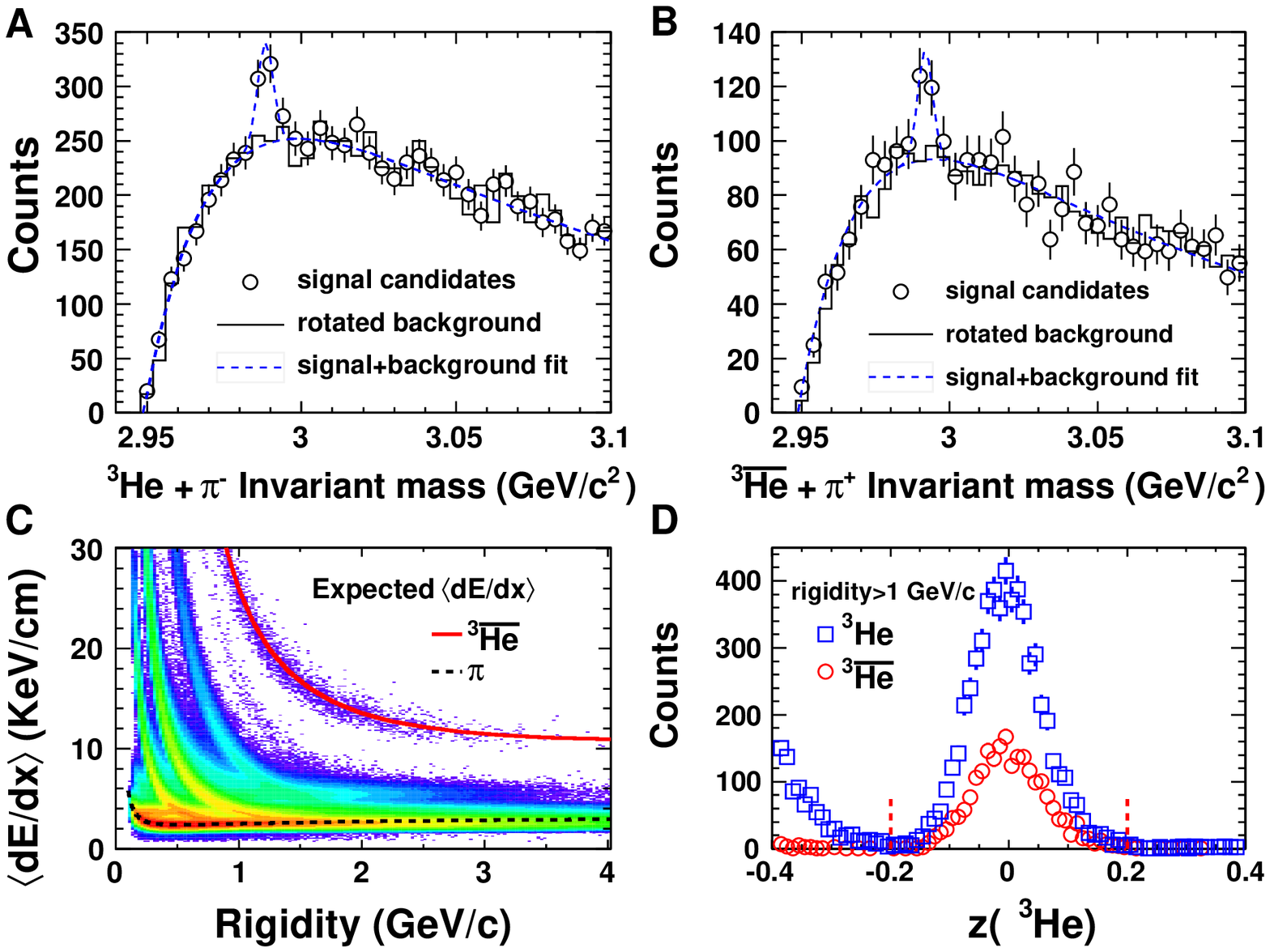,width=3.2in}\vspace{1mm}}
 {\baselineskip
10.5pt\renewcommand{\baselinestretch}{1.05}\footnotesize \noindent
{\bf Fig.~5}\quad ({A} and {B}) Reconstructed invariant mass
distribution of $^{3}He$ and $\pi$, open circles stand for the
signal distribution, while solid lines are the rotated combination
background. Blue dashed lines are the Gaussian (signal) plus
double exponential (background) function fit to the distribution.
({C}) $ \langle dE/dx \rangle$ as a function of rigidity
($p/|Z|$) for negative particles, theoretical $\langle dE/dx
\rangle$ value for \Hebar~ and $\pi$ are also plotted. ({D})
shows that a clean \He~ and \Hebar~ sample can be obtained with
cut $|z(^{3}He)| < 0.2$. Taken from the Ref.~\cite{H3Lbar}.
\label{Invmass} }\vspace{4mm}

\vspace{3mm}
\centerline{\psfig{figure=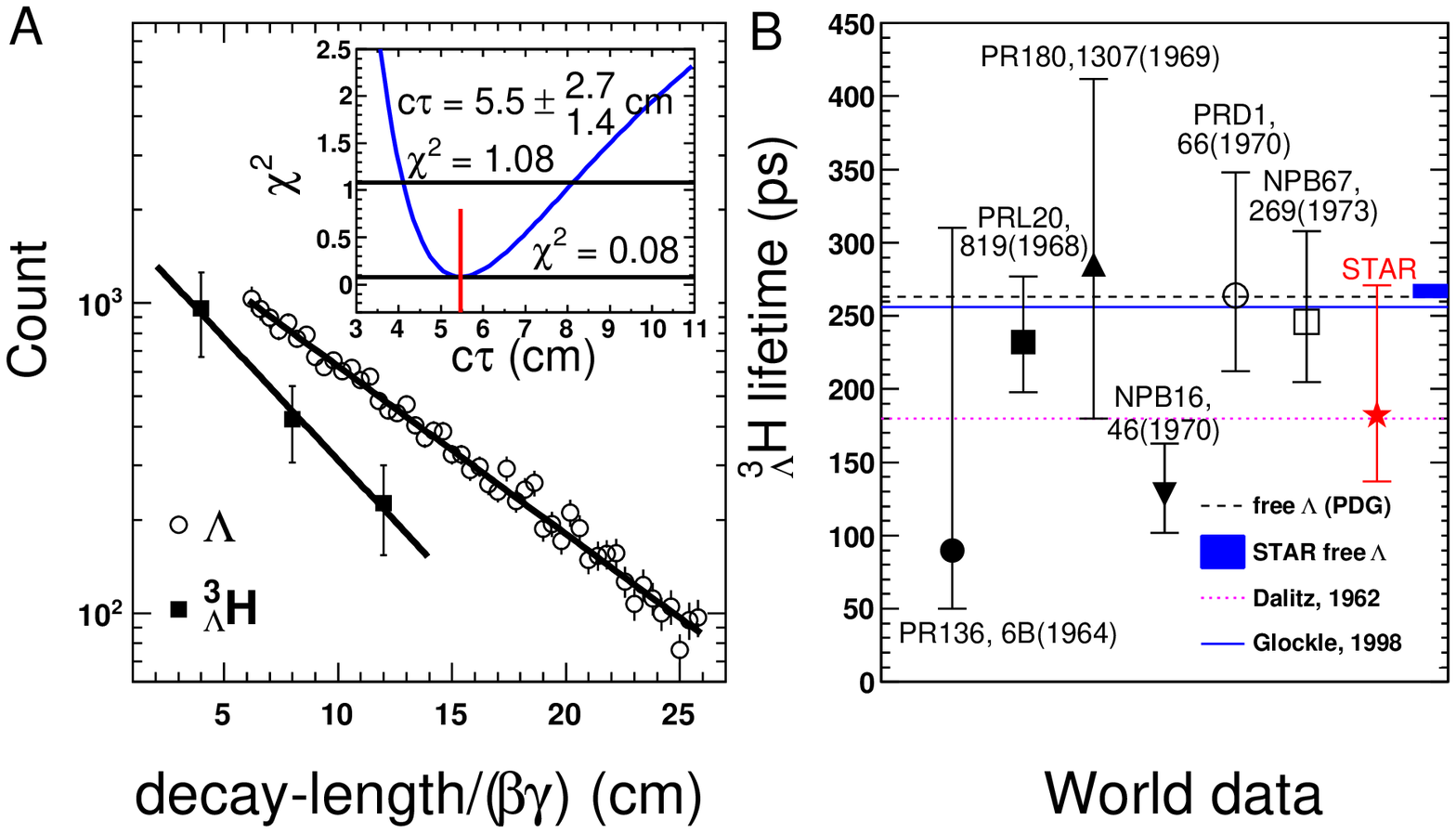,width=3.2in}\vspace{1mm}}
 {\baselineskip
10.5pt\renewcommand{\baselinestretch}{1.05}\footnotesize \noindent
{\bf Fig.~6}\quad ({A}) The yields of \hypertbar~ (solid
squares) and $\Lambda$  (open circles) vs $c\tau$ distribution.
The solid lines stand for the $c\tau$ fits, and the insert plot
describes $\chi^{2}$ distribution of the best fits. ({B})
Comparison between the present measurement and theoretical
calculation  \cite{YN1,YN2}, as well as the previous measurements
\cite{Pre1,Pre2,Pre3,Pre4,Pre5,Pre6}.  Taken from the
Ref.~\cite{H3Lbar}. \label{lifetime} }\vspace{4mm}

The measurement of \hypert~(\hypertbar) lifetime provides us an
effective tool to  understand the Y($\Lambda$)-N(p,n) interactions
\cite{YN1,YN2}.  And, the secondary vertex reconstruction of
\hypert~(\hypertbar) makes us to be able to perform a calculation
of its lifetime, via equation $N(t) = N(0)exp(-t/\tau)$, where $t
= l/(\beta \gamma c)$, $\beta \gamma c = p/m$, $l$ is the decay
length of \hypert, $p$ is their momentum, $m$ is their mass value,
while $c$ is the speed of light. \hypert~and \hypertbar~samples
are combined together to get a better statistics, with the
assumption of the same lifetime of \hypert~ and \hypertbar~ base
on the CPT symmetry theory. The measured yield is corrected with
the tracking efficiency and acceptance of TPC, as well as the
reconstruction efficiency of \hypert~ and \hypertbar. Then, the
$l/(\beta \gamma)$ distribution can be fitted with an exponential
function to extract the lifetime parameter $c\tau$. The best
fitting with $\chi^{2}$ minimization method result in $c\tau =
5.5^{+2.7}_{-1.4} \pm 0.08$, which corresponds a lifetime of
$182^{+89}_{-45} \pm 27$ ps as shown in Figure 6 shows a
comparison between the present measurement and theoretical
calculation  \cite{YN1,YN2}, as well as the previous measurements
\cite{Pre1,Pre2,Pre3,Pre4,Pre5,Pre6}. It seems that the present
measurement of \hypert~ lifetime is consistent with calculation
with phenomenological \hypert~ wave function  \cite{YN1} and a
more recent three-body calculation  \cite{YN2}. The result is also
comparable with the lifetime of free $\Lambda$ within errors.

\vspace{3mm}
\centerline{\psfig{figure=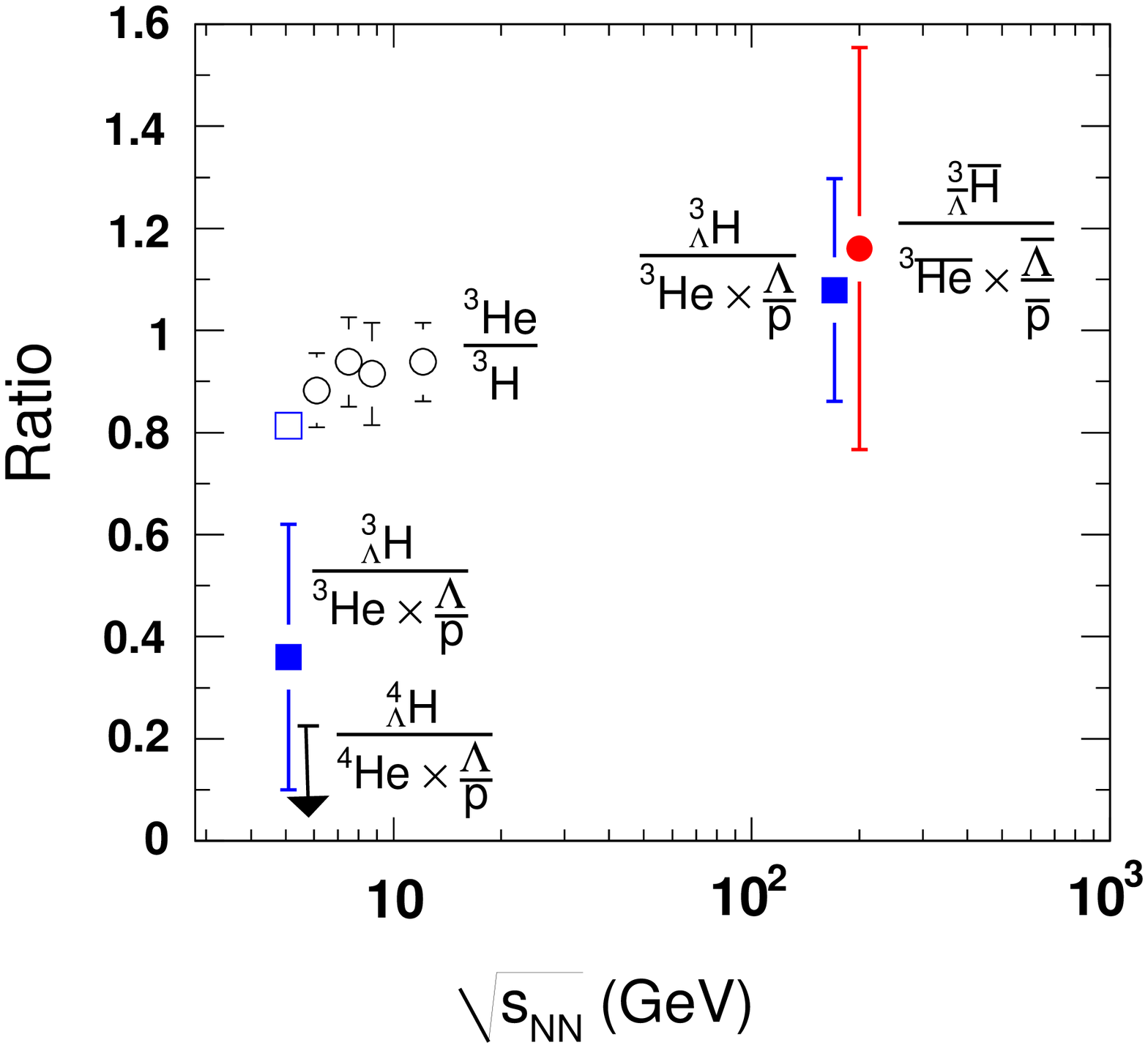,width=3.2in}\vspace{1mm}}
 {\baselineskip
10.5pt\renewcommand{\baselinestretch}{1.05}\footnotesize \noindent
{\bf Fig.~7}\quad  Particle ratios versus center of mass energy
per nucleon-nucleon  collision. The data points besides this
measurement are taken from  Refs.~\cite{Ratio1,Ratio2,Ratio3}.
Only statistical error are presented in the plot.  Taken from the
Ref.~\cite{H3Lbar}. \label{ratio} }\vspace{4mm}


High production rate of \hypert~(\hypertbar) due to equilibration
among strange quarks  and light quarks (u,d) is proposed to be a
signature of the formation of QGP  \cite{H3Lbar,QGP}. The baryon
strangeness correlation factor can be extracted by comparing the
yields of \hypert~ and \He. On the other hand, a recent
calculation  \cite{H3LSong} indicates that the strangeness
population factor, $S_{3} = ^{3}_{\Lambda}H/(
^{3}He\times\Lambda/p)$ is an effect tool to distinguish QGP phase
and pure hadronic phase. Figure 7 depicts the excitation function
of particle ratios for the STAR  data and other previous
measurement~\cite{Ratio1,Ratio2,Ratio3}.  The value of $S_{3}$ is
near unity at RHIC energy, while the same factor is only 1/3 at
AGS energy, indicating that the phase space population of
strangeness are comparable with light quarks at RHIC.

\section{Observation of \Heebar}

The STAR collaboration also reported its observation of
\Heebar~nucleus  \cite{He4bar,Xue}  in April 2011, with 10 billion
gold-gold collisions taken in the year 2007 and 2010. In
additional to the particle identification method by combining
energy loss ($\langle dE/dx \rangle$) and rigidity provided by
TPC, the observation of \Heebar~nucleus relies on the measured
traveling time of tracks given by the barrel TOF  \cite{TOF},
which is composed of 120 trays, surrounding the TPC. With the
barrel TOF, the mass value of particles can be calculated via $m^2
= p^{2}(t^2/L^2 -1)$ (where $t$ and $L$ are the time of flight and
path length of tracks, respectively) for particle identification.
On the other hand, the online high level trigger (HLT) was
employed to perform preferential selection of collisions, which
contains tracks with charge $Ze=\pm2e$ for fast analysis. The
trigger efficiency for \Heebar~is about 70\% with respect to
offline reconstruction, with selection rate less than 0.4\%.
Figure 8 presents the $\langle dE/dx \rangle$ versus rigidity
($p/|Z|$) distribution. The colored bands stand for the helium
sample collected by HLT. A cut of the DCA less than 3 cm for
negative tracks (0.5 cm for positive tracks) is used to reject the
background. In the left panel, 4 \Heebar~candidates are identified
and well separated from \Hebar~at the low momentum region. While a
clear \Hee~signal is presented and centered around the expected
$\langle dE/dx \rangle$ value of \Hee~in the right panel.

\vspace{3mm}
\centerline{\psfig{figure=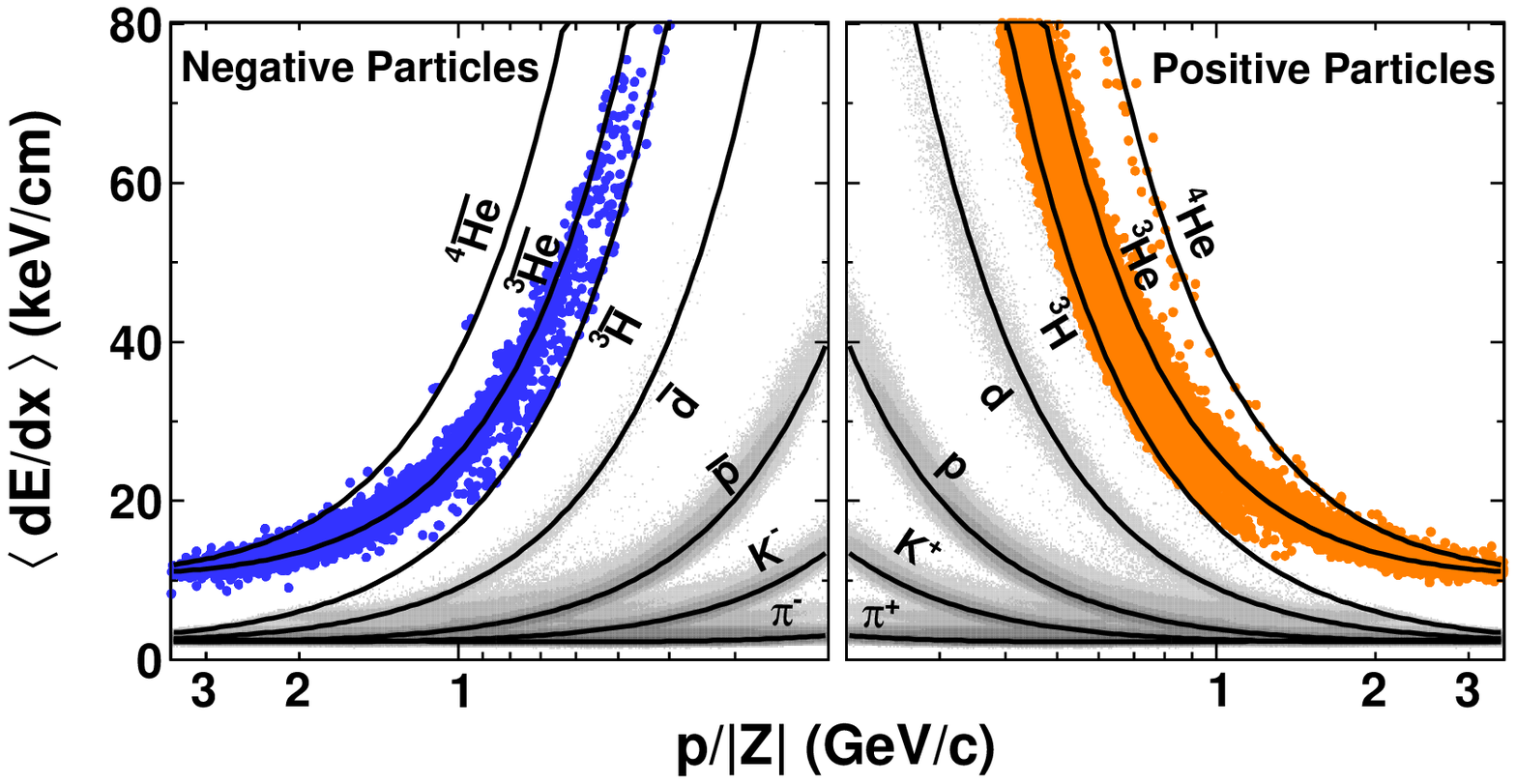,width=3.2in}\vspace{1mm}}
 {\baselineskip
10.5pt\renewcommand{\baselinestretch}{1.05}\footnotesize \noindent
{\bf Fig.~8}\quad  $\langle dE/dx \rangle$ as a function of
$p/|Z|$) for negatively charged  particles (left panel) and
positively charged particles (right panel). The black curves
represent the expected values for each particle species. The lower
edges of the colored bands correspond to the HLT's online
calculation of 3$\sigma$ below the $\langle dE/dx \rangle$ band
center for \He. The grey bands indicate the  $\langle dE/dx
\rangle$ of deuteron, proton, kaon, pion from Minimum bias events
at 200GeV.  Taken from the Ref.~\cite{He4bar}. \label{dEdx}
}\vspace{4mm}

The $\langle dE/dx \rangle$ of \He~(\Hebar) and \Hee~(\Heebar~)
merge together at  higher momentum region, and
$n_{\sigma_{dE/dx}}$, which is defined as $n_{\sigma_{dE/dx}}$ =
$\frac{1}{R}\ln(\langle dE/dx \rangle/\langle dE/dx \rangle^{B} )$
($R$ is the resolution of $\langle dE/dx \rangle$) is used for
further particle identification. Figure 9 shows the combined
particle identification with $n_{\sigma_{dE/dx}}$ and
$mass^{2}/Z^{2}$ value distribution. Two clusters of \Heebar~and
\Hee~located at $n_{\sigma_{dE/dx}} = 0$, $mass^{2}/Z^{2}$ =  3.48
$(GeV/c^{2})^{2}$ can be clearly separated from \Hebar~and \He~as
well as $^3{\rm H}$ and \tbar~are presented in the top panel and
bottom panel. By counting \Heebar~signal with the cuts window $-2
<n_{\sigma_{dE/dx}} < 3$ and 2.82 $(GeV/c^{2})^{2}$ $<
mass^{2}/Z^{2} <$ 4.08 $(GeV/c^{2})^{2}$ as indicated in the top
panel, 16 \Heebar~candidates are identified. Together with 2
\Heebar~candidates detected by TPC alone in the year 2007 which is
presented in the figure, 18 \Heebar~candidates are observed by the
STAR experiment. So far, \Heebar~ is the heaviest antimatter
nucleus observed in the world. Right after the public report of
\Heebar~ from the STAR collaboration, the LHC-ALICE collaboration
also claimed the observation of 4 \Heebar~ particles
\cite{ALICE2}.

\vspace{3mm}
\centerline{\psfig{figure=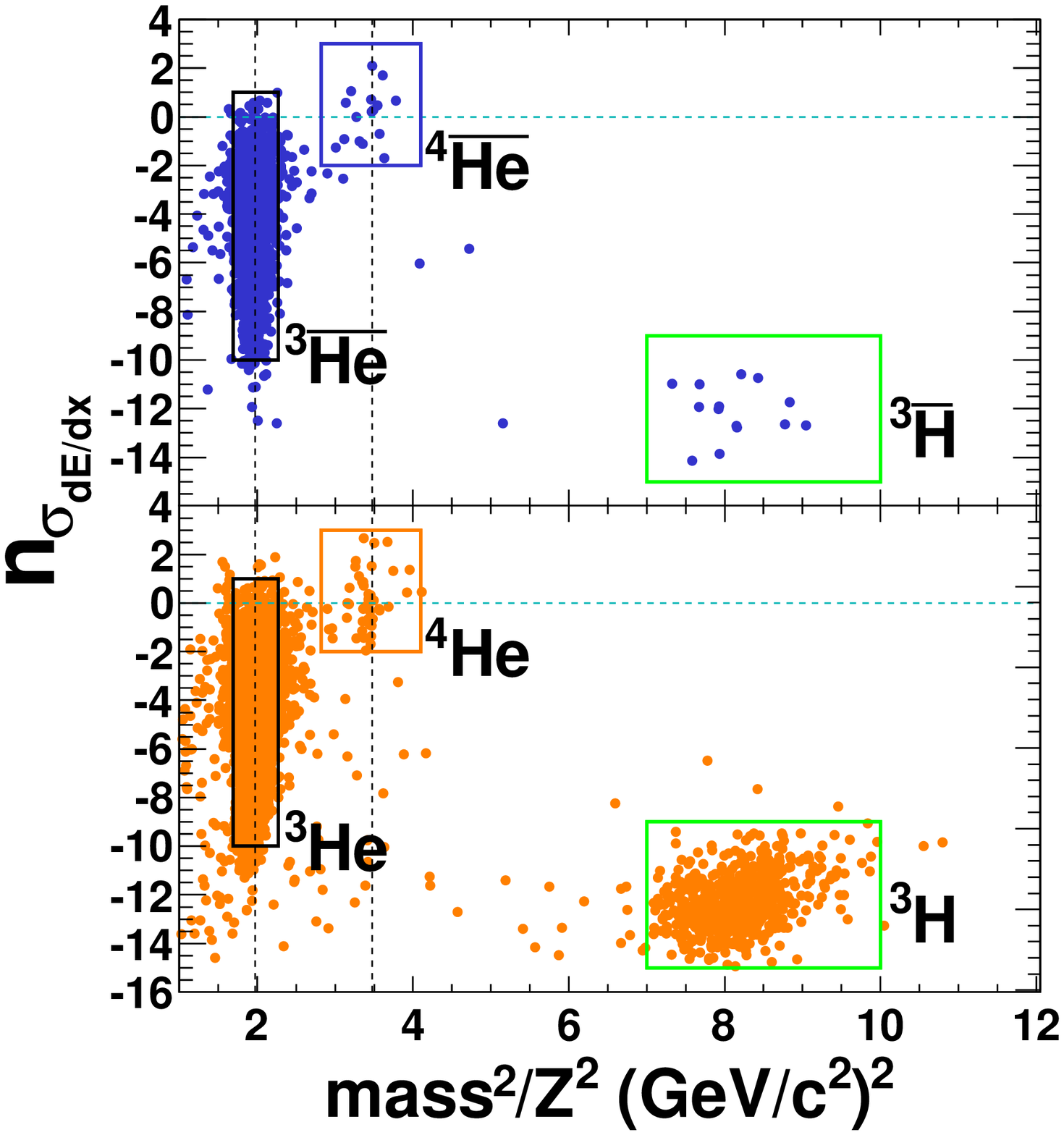,width=3.2in}\vspace{1mm}}
 {\baselineskip
10.5pt\renewcommand{\baselinestretch}{1.05}\footnotesize \noindent
{\bf Fig.~9}\quad  Top (bottom) panel shows the
$n_{\sigma_{dE/dx}}$ versus  $mass^{2}/Z^{2}$ distribution for
negative (positive) particles. The horizontal dashed lines mark
the $n_{\sigma_{dE/dx}} = 0$, while the vertical ones stand for
the theoretical mass values of \He(\Hebar) and \Hee(\Heebar). The
signals of \Heebar~and \Hee~are counted in the cuts window of
$-2<n_{\sigma_{dE/dx}} <3.$ and $2.82 (GeV/c^{2})^{2} <
mass^{2}/Z^{2} < 4.08 (GeV/c^{2})^{2}$. \label{tofpid}
}\vspace{4mm}


\section{Effort to search for antinuclei in Cosmic rays}

As we discussed in above sections, most efforts on hunting
antinuclei were realized in high-energy nuclear physics
laboratories. Nevertheless, it is still a dream to capture any
antinucleus in cosmos. The search of \Heebar~ and heavier
antinucleus in universe is one of the major motivations of space
shuttle based apparatus such as the Alpha Magnetic Spectrometer
\cite{AMS}, both the RHIC-STAR experimental result and model
calculation provide a background estimation of \Heebar~ for the
future observation in Cosmos radiations \cite{He4bar}. Recently,
the effort to search for the Cosmic-Ray Antideuterons and
Antihelium by the Balloon-borne Experiment with Superconducting
Spectrometer (BESS) collaboration has been made
\cite{BESS1,BESS2}. However, neither  Antideuterons candidate was
found using data collected during four BESS balloon flights from
1997 to 2000 \cite{BESS1}, nor  Antihelium candidate was found in
BESS-Polar I data among 8.4 $\times$ $10^6$ $|Z|$ = 2 nuclei from
1.0 to 20 GV (absolute rigidity) or in BESS-Polar II data among
4.0 $\times$ $10^7$ $|Z|$=  2 nuclei from 1.0 to 14 GV
\cite{BESS2}. They derived an upper limit of 1.9 $\times$
$10^{-4}$ ($m^2 ~s ~sr GeV/nucleon)^{-1}$ for the differential
flux of cosmic-ray antideuterons, at the 95$\%$ confidence level,
between 0.17 and 1.15 GeV/nucleon at the top of the atmosphere
\cite{BESS1}. For antihelium, assuming antihelium to have the same
spectral shape as helium, a 95$\%$ confidence upper limit to the
possible abundance of antihelium relative to helium of 6.9
$\times$ $10^{-8}$ was determined combining all BESS data,
including the two BESS-Polar flights.  With no assumed antihelium
spectrum and a weighted average of the lowest antihelium
efficiencies for each flight, an upper limit of 1.0 $\times$
$10^{-7}$ from 1.6 to 14 GV was determined for the combined
BESS-Polar data. Under both antihelium spectral assumptions, these
are the lowest limits obtained to date \cite{BESS1}. Fig. 12 shows
the new upper limits of antihelium/helium from the BESS experiment
\cite{BESS1}. It seems very difficult to hunt anti-helium in
cosmos.

\vspace{3mm}
\centerline{\psfig{figure=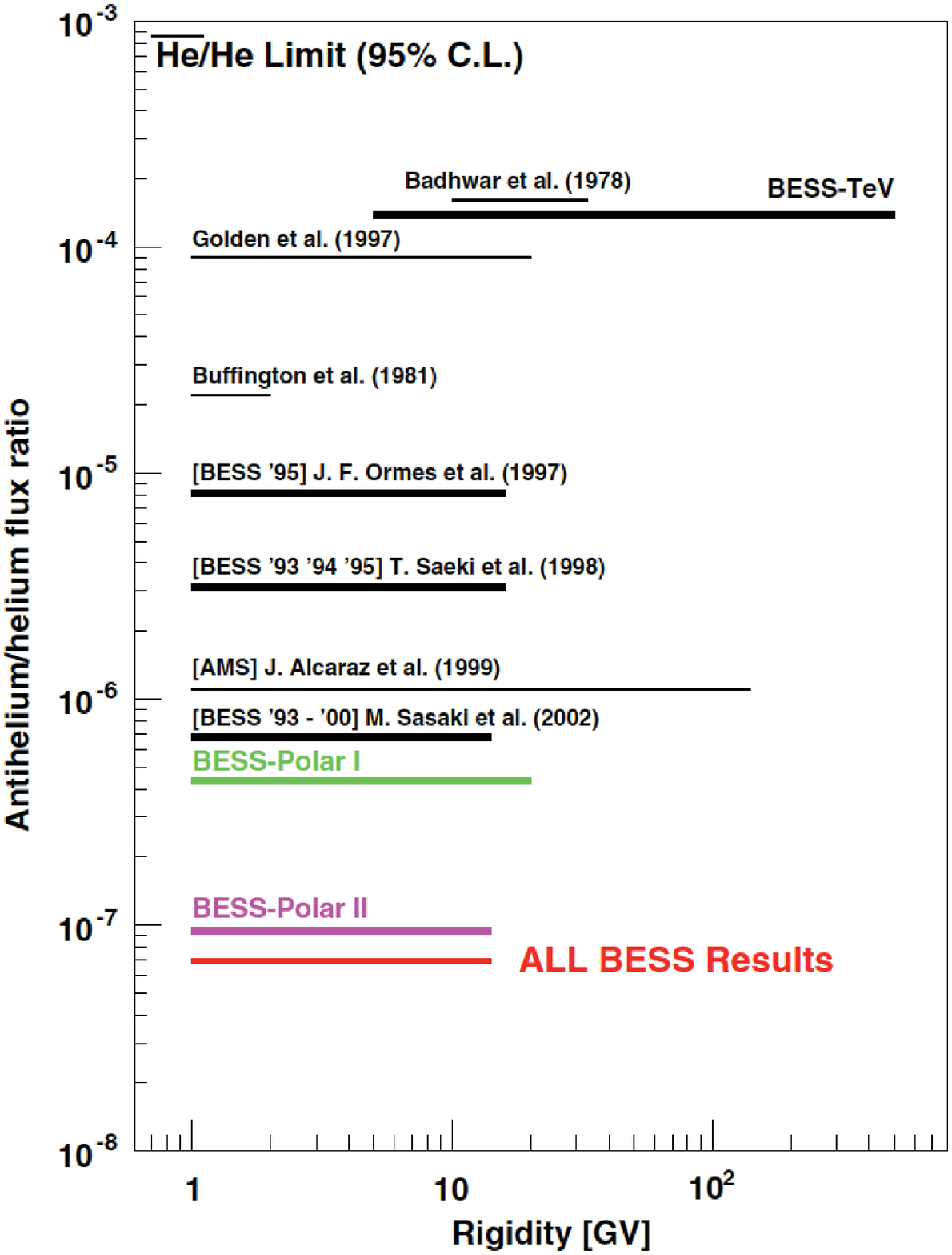,width=3.2in}\vspace{1mm}}
 {\baselineskip
10.5pt\renewcommand{\baselinestretch}{1.05}\footnotesize \noindent
{\bf Fig.~10}\quad The new upper limits of antihelium/helium at the
TOA calculated assuming the same energy spectrum for He as
for He with previous experimental results. The limit
calculated with no spectral assumption is about 25$\%$ higher.
Taken from Ref.~\cite{BESS2}.
\label{yield}
}\vspace{4mm}

\section{Trap of antihydrogen atoms}

Parallel to huge efforts of searching for the antimatter nuclei in
the  cosmos and laboratory, some scientists are thinking how to
build the antimatter atom and make it trapped. Antihydrogen, the
simplest form of antiatoms, which is the bound state of an
antiproton and a positron, was reported to be produced by the LEAR
in 1996  \cite{AntiH1st} and at low energies at CERN (the European
Organization for Nuclear Research) since 2002
\cite{Anti-H1,Anti-H2}. Antihydrogen is of interest for use in a
precision test of nature's fundamental symmetries which is one of
the most important projects of experiments in physics. In the
standard model of particle physics, all properties of a particular
physical process should be identical under the operation of charge
conjugation, parity reflection, and time reversal (CPT). In
nuclear physics, all properties of hydrogen including the fine
structure and hyperfine structure are known at a high precision. A
precise comparison between antihydrogen and hydrogen spectrum is
one the best ways to measure the possible CPT violation. The first
confinement of antihydrogen with a trap time of 172 ms was
demonstrated by the ALPHA experiment in 2010 \cite{Anti1trap}
based on the antiproton decelerator facility. The most recent
results with a confine time of 1000s was present by the same
collaboration  \cite{Anti2trap}, which is a big step towards the
measurement of the antihydrogen properties. On the other hand, the
mutual repulsive force between antimatter and matter
(anti-gravity) was proposed, and many theoretical works have been
done  \cite{AntiG,AntiG1} since the first observation of
antimatter. However, no experimental result can be obtained until
now. The successful trap of antihydrogen atoms provides a method
to measure the gravitational effects by reducing the antihydrogen
temperatures to the sub mK level in the way of adiabatic cooling.

\section{ Production mechanisms of antimatter light nuclei}\vspace{2.5mm}

\noindent

Antimatter particles including
\ebar~,\pbar~,\db~,\Hebar~,\hypertbar~,\Heebar~and  antihydrogen
atoms were observed in the past eighty years based on different
kinds of sources and detectors. Most of these antimatter particles
were produced by nucleon-nucleon reactions, where their production
rate can be described by both thermodynamic model and coalescence
model. In thermodynamic model, the system created is characterized
by the chemical freeze-out temperature ($T_{ch}$), kinetic
freeze-out temperature ($T_{kin}$), as well as the baryon and
strangeness chemical potential $\mu_{B}$ and $\mu_{S}$,
respectively. (Anti)nucleus is regarded as an object with energy
$E_{A}=Am_{p}$ (A is the atomic mass number, $m_{p}$ is the mass
of proton) emitted by the fireball  \cite{PBM}. The production
rate are proportional to the Boltzmann factor $e^{-m_{p}A/T}$ as
shown in Equ.~(\ref{eq:thermal}), where $P_{A}$ and $g$ are the
momentum and degeneracy of (anti)nucleus, $V$ is the volume of the
fireball. In coalescence picture, (anti)nucleus is formed by
coalescence at the last stage of the system evolution since there
exists strong correlation between the constituent nucleons in
their phase space  \cite{STARdHe3,Sato,Scheibl1999}. The
production probability is described by
Equ.~(\ref{eq:Coalescence}),  where $E\frac{d^{3}N}{d^{3}p}$
stands for the invariant yield of nucleons or (anti)nucleus, $Z$
is the atomic number. And, $p_{A},p_{p},p_{n}$ are the momentum of
(anti)nucleus, protons and neutrons, with $p_{A} = A\times p_{p}$
is assumed. $B_{A}$ is the coalescence parameter.

\vspace{3mm}
\centerline{\psfig{figure=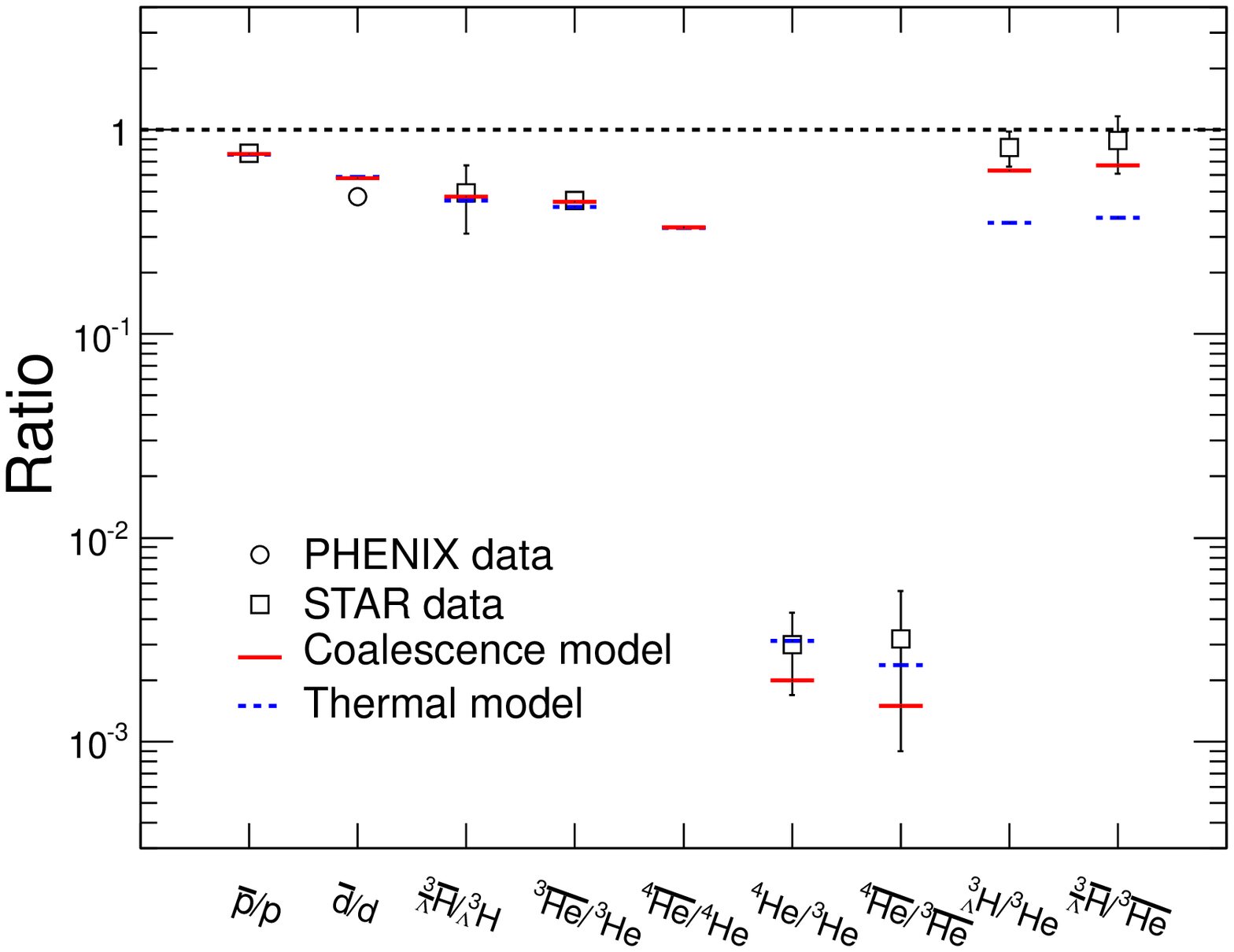,width=3.2in}\vspace{1mm}}
 {\baselineskip
10.5pt\renewcommand{\baselinestretch}{1.05}\footnotesize \noindent
{\bf Fig.~11}\quad  The comparison of particle ratios between data
and model calculations.  The data points are taken from the STAR
and the PHENIX experiments
\cite{H3Lbar,He4bar,Phenixdbar,ppbarratio}. The coalescent results
are based on naive coalescence algorithm with a momentum
difference lower than 100MeV and a coordinator space difference
less than 2R (R is the nuclear force radius), while the thermal
predication is taken from   \cite{PBM}. Taken from
Ref.~\cite{Xue-Ma}.

\label{ParticleRatios}
}\vspace{4mm}

\begin{equation}
E_{A}\frac{d^{3}N_{A}}{d^{3}P_{A}} =  \frac{gV}{(2\pi)^{3}}E_{A}e^{-m_{p}A/T},\\\
\label{eq:thermal}
\end{equation}

\begin{equation}
E_{A}\frac{d^{3}N_{A}}{d^{3}P_{A}} = B_{A}(E_{p}\frac{d^{3}N_{p}}{d^{3}P_{p}})^{Z}(E_{n}\frac{d^{3}N_{n}}{d^{3}P_{n}})^{A-Z}.\\\
\label{eq:Coalescence}
\end{equation}
Relative particle production abundance of (anti)nucleus are
explored based on thermal model  and coalescence model, and
compared with data taken at RHIC. Figure 11 shows the particle
ratios of (anti)nucleus, both thermal model and coalescence model
can fit the antinucleus to nucleus ratios at RHIC energy. While
the coalescence model has a better description for \hypert/\He~and
\hypertbar/\Hebar~than thermal model \cite{Xue-Ma}. In a
microscopic picture, both coalescent and thermal production of
(anti)nucleus predict an exponential trend for the production rate
as a function of baryon number. The exponential behavior of
(anti)nucleus production rate in nuclear nuclear reaction has been
manifested in Figure 12, which depicts the invariant yields
($d^{2}N/(2\pi p_{T}dp_{T}dy)$) evaluated at the average
transverse momentum ($p_{T}/|B| = 0.875GeV/c$) region versus
baryon number distribution. The solid symbols represent results
reproduced by the coalescence model, which fits the data points
very well. By fitting the model calculation with an exponential
function $e^{-r|B|}$, a reduction rate of 1692 (1285) can be
obtained for each additional antinucleon (nucleon) added to
antinucleus (nucleus), compared to $1.6^{+1.0}_{-0.6}\times10^{3}$
($1.1^{+0.3}_{-0.2}\times10^{3}$) for nucleus and (antinucleus)
obtained by the STAR experiment. The yield of next stable
antinucleus (antilithium-6) is predicted to be reduce by a factor
of $2.6\times10^{6}$ compare to \Heebar~, and is impossible to be
produced within current accelerator technology. The excitation of
(anti)nucleus from a highly correlated vacuum was discussed in
reference   \cite{newphysics}. This new production mechanism can
be tested with the measurement of the production rate of
(anti)nucleus, any deviation of the production rate of
(anti)nucleus from usual reduction rate, may indicate the exist of
the direct excitation mechanism. On the other hand, the low
production rate of \Heebar~ antinucleus created by nuclear
interaction indicate that any observation of \Heebar~ or even
heavier antinucleus should be a great hint of the existence of
large amount of antimatter somewhere in the Universe.

\vspace{3mm}
\centerline{\psfig{figure=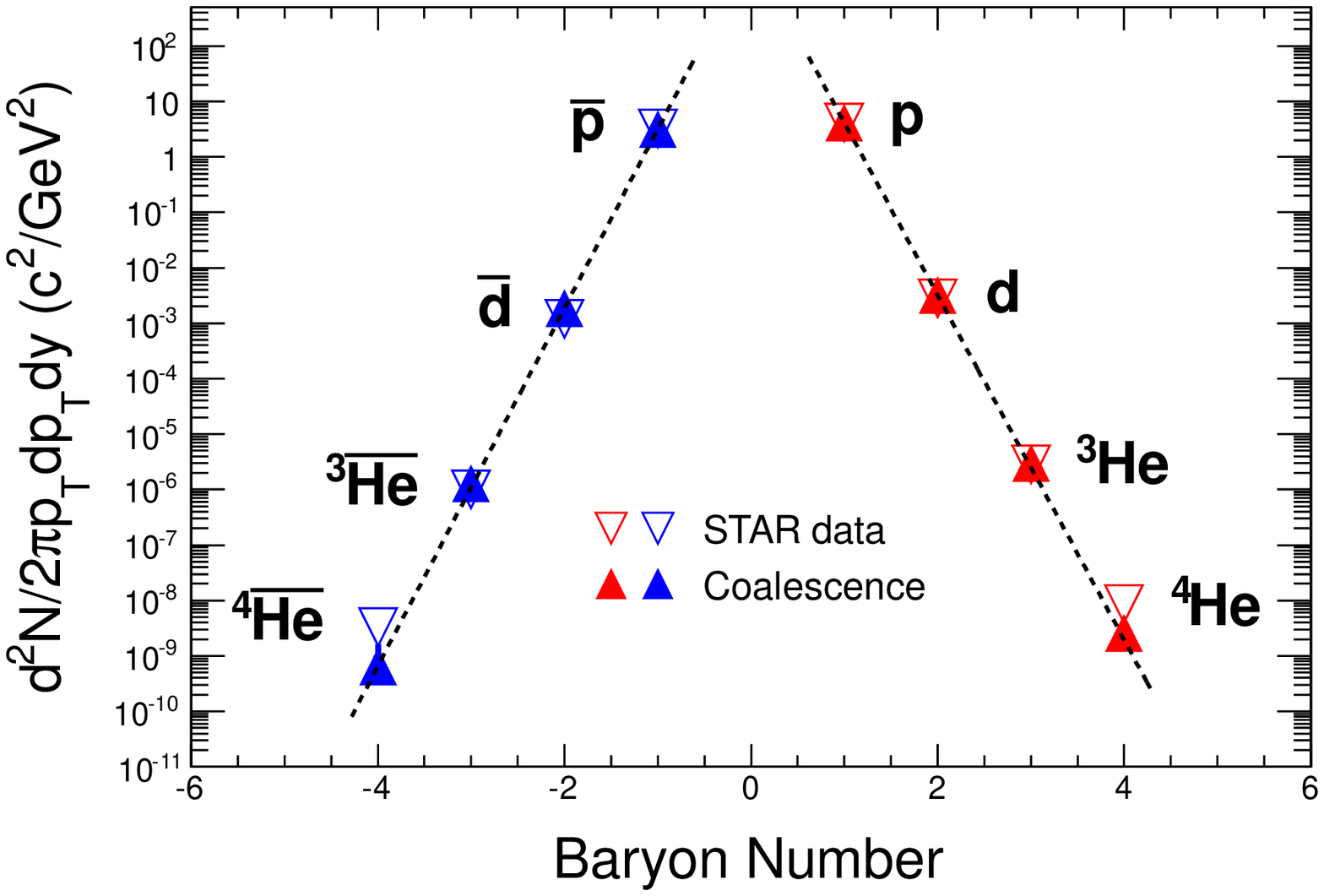,width=3.2in}\vspace{1mm}}
 {\baselineskip
10.5pt\renewcommand{\baselinestretch}{1.05}\footnotesize \noindent
{\bf Fig.~12}\quad Invariant yields $d^{2}N/(2\pi p_{T}dp_{T}dy)$
of (anti)nucleus at  the average transverse momentum region
($p_{T}/|B| = 0.875GeV/c$) as a function of baryon number (B). The
open symbols represents the data points extracted by the STAR
experiment at RHIC energy, while solid ones are reproduced by
coalescence model.  The lines represent the exponential fit for
our coalescence results of positive particles (right) and negative
particles (left) with formula $e^{-r|B|}$. Taken from
Ref.~\cite{Xue-Ma}. \label{yield} }\vspace{4mm}

\section{Conclusions and Perspectives}\vspace{2.5mm}

We briefly review the discovery history of antimatter, including
positron which is the  first antimatter observed
\cite{Chao2,Positron}, antihelium-4 which is the heaviest
antimatter nucleus observed so far \cite{He4bar} as well as
anti-hypertriton which is the first antimatter hypernucleus
\cite{H3Lbar}.
With the increasing of mass of antimatter particles, the
difficulty of observation  becomes much larger. With the increase
of accelerator  technology and beam energy, the detection of
heavier antimatter nuclei becomes feasible. Anti-hypertriton and
antihelium-4 are two good examples for antimatter detection with
the present relativistic heavy ion collision facility. In the
viewpoint of antimatter production, thermal model and coalescence
model can basically describe the production yield of antimatter
and antimatter-matter ratio. In our recent calculation based on
the hydrodynamic motivated BlastWave model couple with a
coalescence model at RHIC energy, we demonstrate that the current
approach can reproduce the differential invariant yields and
relative production abundances of light antinuclei and
antihypernuclei \cite{Xue-Ma}. The exponential behavior of the
differential invariant yields versus baryon number distribution is
studied. By extrapolating the distribution to B = -6 region, the
production rate of \Libar~ in high energy heavy ion collisions is
about $10^{-16}$, and seems  impracticably to be observed within
current accelerator technology. As addressed in Sec. 5, the
observation of \Heebar~ and even heavier antinuclei in Cosmic rays
is a great hint of the existence of massive antimatter in
Universe. The model calculations and experimental measurements in
high energy heavy ion collisions can simulate the interactions
between high energy protons and interstellar materials. Thus our
results provide a good background estimation for the future
observation of \Heebar~and even heavier antinuclei in Universe.

\noindent {\baselineskip
10.5pt\renewcommand{\baselinestretch}{1.05}\footnotesize {\bf
Acknowledgements}~~~This work is partially supported by the NSFC
under  contracts No. 11035009, 11220101005, 10875160, 10805067 and
10975174, 11275250 and 10905085, the Knowledge Innovation Project
of Chinese Academy of Sciences under Grant No. KJCX2-EW-N01, and
Dr. J. H. Chen is partially supported by the Shanghai Rising Star
Project under Grand No. 11QA1408000. }\vspace{7mm}
\hrule\vspace{4mm}

\noindent {\large
\usefont{T1}{fradmcn}{m}{n}\bf References}\vspace{3.1mm}

\parskip=0mm \baselineskip 15pt\renewcommand{\baselinestretch}{1.25} \footnotesize
\parindent=9mm

\setlength{\baselineskip}{12.3pt}   

\end{document}